%% file: main.tex
\documentclass[conference]{IEEEtran}

\usepackage{cite}
\usepackage{amsmath,amssymb,amsfonts}
\usepackage{algorithmic}
\usepackage{graphicx}
\usepackage{subcaption}
\usepackage{float}
\usepackage{stfloats}
\usepackage{textcomp}
\usepackage{xcolor}
\usepackage{booktabs}
\usepackage{multirow}
\usepackage{tabularx}
\usepackage{array}
\usepackage{xurl}
\usepackage{enumitem}
\usepackage{hyperref}
\usepackage{makecell}
\usepackage[table]{xcolor}
\usepackage[most]{tcolorbox}
\hypersetup{hidelinks}

\definecolor{promptbg}{RGB}{248,248,248}
\definecolor{promptframe}{RGB}{120,120,120}
\newtcblisting{promptbox}[2][]{%
  enhanced,
  breakable,
  colback=promptbg,
  colframe=promptframe,
  boxrule=1pt,
  arc=1pt,
  left=1pt,
  right=1pt,
  top=1pt,
  bottom=1pt,
  boxsep=1pt,
  toptitle=1pt,
  bottomtitle=1pt,
  before skip=1pt,
  after skip=1pt,
  listing only,
  listing options={%
    basicstyle=\ttfamily\footnotesize,
    breaklines=true,
    breakatwhitespace=true,
    breakautoindent=false,
    breakindent=0pt,
    columns=fullflexible,
    keepspaces=true,
    showstringspaces=false,
    xleftmargin=0pt,
    aboveskip=1pt,
    belowskip=1pt,
    lineskip=1pt
  },
  title={#2},
  fonttitle=\bfseries\scriptsize,
  #1
}

\def\BibTeX{{\rm B\kern-.05em{\sc i\kern-.025em b}\kern-.08em
    T\kern-.1667em\lower.7ex\hbox{E}\kern-.125emX}}

\IEEEoverridecommandlockouts

\makeatletter
\def\@IEEEpubidpullup{3.0\baselineskip}
\makeatother

\begin{document}
\bstctlcite{BSTcontrol}

\title{From Retrieved Evidence to Security Outcomes: \\ A Data-Driven Analysis of Java Security API Misuse in LLM-Generated Code
}

\author{\IEEEauthorblockN{Tianhe Lu, Eric Spero, Sakuna Harinda Jayasundara, Robert Biddle, Giovanni Russello}
\IEEEauthorblockA{\textit{School of Computer Science} \\
\textit{University of Auckland}\\
Auckland, New Zealand \\
tlu202@aucklanduni.ac.nz, eric.spero@auckland.ac.nz, sjay950@aucklanduni.ac.nz, \\
robert.biddle@auckland.ac.nz,
g.russello@auckland.ac.nz}}

\IEEEpubid{%
  \parbox[b]{\textwidth}{%
    \fontsize{7.5}{8.5}\selectfont
    \raggedright
    \textcopyright~2026 IEEE.
    Personal use of this material is permitted.
    Permission from IEEE must be obtained for all other uses,
    in any current or future media, including reprinting/republishing
    this material for advertising or promotional purposes, creating
    new collective works, for resale or redistribution to servers
    or lists, or reuse of any copyrighted component of this work
    in other works.
  }%
}

\maketitle

\maketitle

\begin{abstract}
The misuse of Java security APIs is a serious security problem in software development. Large language models can generate compilable code, while still misusing cryptographic and secure-communication APIs. We investigate whether external security evidence can reduce this risk and how retriever choice and evidence source shape the resulting programs. Our study evaluates the GPT-5.5 model and Llama-3.3-70B-Instruct model on 12 Java Cryptography Architecture (JCA) and Java Secure Socket Extension (JSSE) security functionalities under 11 prompt and evidence conditions, yielding 23,760 generated programs. We measure program validity, detected Java security API misuse, and the top-$5$ evidence selected by four full-corpus retrievers. Under task-only prompting, 58.5\% of valid Llama-3.3-70B-Instruct-generated programs and 36.6\% of valid GPT-5.5-generated programs contain detected security API misuse. 
Dense full-corpus retrieval reduces conditional misuse among valid programs by 29.0--32.7 and 22.4--31.0 percentage points, respectively. The retrievers show preferences for substantially different evidence types. BM25 and the OpenAI embedding retrievers favor official guidance, the Qwen embedding retriever favors code examples, and the Jina embedding retriever selects a more heterogeneous mixture. Under source-restricted retrieval, code examples yield the lowest observed conditional misuse for the Llama-3.3-70B-Instruct model, while explicit misuse patterns produce zero detected misuse cases for known benchmark patterns among 1,016 valid programs generated by the GPT-5.5 model, with validity reduced to 94.1\%. External evidence can substantially reduce detected misuse in the evaluated settings, and the results motivate actionable evidence selection and post-generation validation. 
\end{abstract}


\section{Introduction}  \label{sec:Introduction}

Large Language Models (LLMs) have been deeply integrated into software development workflows. 
Code completion, function generation, and repository-level development assistance are increasingly used in real-world software engineering practice \cite{noauthor_2025_nodate, jiang_survey_2024, chen_evaluating_2021}. 
However, together with the productivity gains, concerns about the security of LLM-generated code continue to grow \cite{mousavi_investigation_2024, pearce_asleep_2025, mohsin_can_2024,  khoury_how_2023, zhuo_identifying_2026, zhang_llm_2025, fu_security_2025, dou_whats_2025}. 
Existing research has repeatedly demonstrated that LLM-generated code is not naturally secure and may reproduce insecure patterns present in training data \cite{pearce_asleep_2025, khoury_how_2023}. 

\IEEEpubidadjcol

This problem is particularly serious in the context of security API usage. 
Security APIs are not simple library interfaces that can be used without careful consideration.
Their correct use often requires programs to satisfy multiple constraints, such as algorithm selection, parameter configuration, call order, secret handling, and identity verification.
The security of using those APIs often depends on whether these details are correct \cite{mosavi_detecting_2025, kruger_crysl_2021}. 
As a result, security API misuse has been a persistent problem in real software ecosystems \cite{egele_empirical_2013, wickert_fix_2022}. 
When LLMs are trained on publicly available code, they may learn both correct usages and misuse patterns, and may output these unsafe continuations during generation \cite{pearce_asleep_2025}.

Mousavi et al. \cite{mousavi_investigation_2024} provided one of the most systematic characterizations of this problem. 
They designed coding tasks for commonly used Java security APIs and evaluated ChatGPT-generated programs through a combination of automated analysis and manual review. 
The results were alarming: around 70\% of the generated code samples contained security API misuse, and some task descriptions reached misuse rates of 100\%. 
Their Java Cryptography Architecture (JCA) and Java Secure Socket
Extension (JSSE) benchmark provides a valuable and controlled testbed for studying whether misuse persists in newer LLM settings and whether external security evidence can change measured outcomes.

One natural intervention is to supply external knowledge. 
Prior work has shown that Retrieval-Augmented Generation (RAG) and in-context examples can improve API-oriented code generation \cite{lewis_retrieval-augmented_2020, brown_language_2020, chen_benchmarking_2024, chen_when_2025}. 
However, most prior evaluations focus on functional correctness or the use of unfamiliar APIs.
JCA and JSSE present a different setting, as a model may already know these APIs while having internalized both secure and insecure usage patterns. 
In this setting, external evidence can do more than supply missing API facts, as it can make implicit security constraints explicit. 
However, it remains unclear how different retrievers and evidence sources affect actual security API misuse in the generated code.

\begin{figure*}[t]
    \centering
    \includegraphics[trim=0 20 0 0,clip,width=\textwidth]{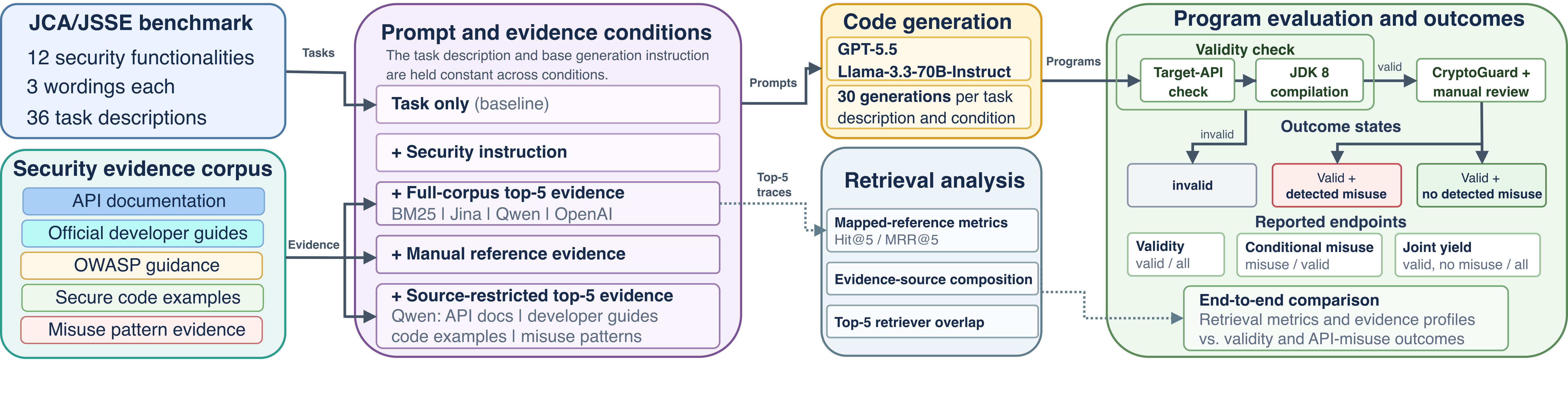}
    \caption{Overview of the experimental and analysis workflow.
    }
    \label{fig:overview}
\end{figure*}

This paper studies this question using the JCA and JSSE benchmark with a heterogeneous security-relevant corpus containing API documentation, official developer guides, OWASP guidance, secure code examples, and misuse patterns. 
Figure \ref{fig:overview} shows the overall workflow. 
We evaluate the GPT-5.5 and Llama-3.3-70B-Instruct models under multiple prompt and evidence conditions, including a task-only baseline, a prompt-only security instruction, four full-corpus retrievers, manually selected reference evidence, and four source-restricted retrieval conditions. 
Generated programs first undergo validity checks, and valid programs are then examined for security API misuse through CryptoGuard \cite{rahaman_cryptoguard_2019} and manual review. 
For automatic retrieval, we analyze the top-$5$ ranked retrieval traces using mapped-reference metrics and evidence-source composition, and compare these retrieval profiles with validity and misuse outcomes.

The paper addresses the following research questions:
\begin{itemize}[noitemsep]
    \item \textbf{RQ1}: How prevalent and task-dependent is detected Java security API misuse in code generated by the evaluated LLMs?
    \item \textbf{RQ2}: How do prompt-only guidance, full-corpus retrieval, and manually selected reference evidence change validity and detected API misuse?
    \item \textbf{RQ3}: How do retriever choice and evidence source shape retrieval profiles and security outcomes?
\end{itemize}

The results show that misuse remains common and concentrated in tasks involving secret handling, parameter selection, and verification logic. 
Dense retrieval substantially reduces detected misuse for both models.
The retrievers construct different evidence selections, and the source-restricted conditions suggest that executable examples and explicit negative constraints provide complementary forms of support. 
These observations motivate security-oriented retrieval that combines actionable evidence with post-generation validation. 

This work makes the following contributions:
\begin{itemize}
    \item We provide a misuse-centered evaluation of two LLM settings across 12 Java cryptographic and secure-communication functionalities and 23,760 generated programs.
    \item We quantify the effects of prompt-only, automatically retrieved, manually selected, and source-restricted security evidence on program validity and detected misuse.
    \item We characterize the evidence selection by different retrievers and relate its source composition to code security outcomes. 
\end{itemize}

\section{Background and Related Work} \label{sec:Related Work}

\subsection{Security of LLM-Generated Code and API Misuse} \label{sec: Security of LLM-Generated Code and API Misuse}
Prior work has shown that syntactically valid or functional LLM-generated code may contain substantial security weaknesses \cite{jiang_survey_2024, pearce_asleep_2025, khoury_how_2023, zhuo_identifying_2026, zhang_llm_2025, fu_security_2025, dou_whats_2025}. 
Security API misuse is a more specific failure mode involving API selection, call sequences, or surrounding validation logic \cite{mosavi_detecting_2025}.
Mousavi et al. reported widespread misuse in ChatGPT-generated Java security code \cite{mousavi_investigation_2024}, while Zhuo et al. identified similar API misuse in other programming languages and API ecosystems \cite{zhuo_identifying_2026}. 
Together, these studies establish security API misuse as a distinct and persistent failure mode in LLM-generated code.

\subsection{Interventions for Secure Code Generation} \label{sec:Interventions for Secure Code Generation}
Secure code generation interventions operate through prompting, model adaptation, decoding, and inference-time evidence. 
Security-focused prompting and security charters provide lightweight guidance \cite{tony_prompting_2025, shukla_security_charter_2025}. 
SafeCoder and CoSec modify model behavior through instruction tuning or supervised co-decoding \cite{he_instruction_2024, li_cosec_2024}.
Retrieval-based approaches supply secure code demonstrations, task-specific guidelines, or vulnerability-oriented knowledge when LLMs generate programs \cite{zhang_seccoder_2024, tony_retrieve_2025, lin_give_2025, shi_rescue_2026}. 
These studies typically treat prompting, fine-tuning, decoding, or retrieval as mechanisms for improving generated code. 
In contrast, we use prompting and retrieval as controlled conditions to characterize how these conditions change validity and detected API misuse to relate their effects to the evidence selected by the retrievers.

\subsection{Retrieval-Augmented Code Generation} \label{sec:Retrieval-Augmented Code Generation}

Retrieval-Augmented Generation (RAG) \cite{lewis_retrieval-augmented_2020} was originally proposed to fill knowledge gaps in LLMs using external knowledge. 
In code generation, it has also been applied to programming problems, repository contexts, and API-oriented generation \cite{chen_when_2025, wang_coderag-bench_2025, lin_give_2025}.
Regarding API-oriented code generation, existing research shows that for less common API libraries, retrieved API documentation can aid LLM code generation, and code examples are generally more helpful than descriptive text \cite{gu_what_2025, chen_when_2025}. 
Further study shows that different LLMs perform differently on different API tasks, and the benefits of RAG are also not the same across different models and tasks \cite{wu_comprehensive_2024}. 
However, these studies primarily focus on the scenario where the LLM is unfamiliar with the API, while our study investigates a different situation where LLM is already familiar with the API but may have learned unsafe use patterns. 
Accordingly, our focus is not merely knowledge supplementation, but whether retrieval can correct unsafe generation biases for APIs the model already ``knows", and how the retrieval profiles relate to validity and detected security API misuse.

\section{Study Design} \label{sec:Study Design}

Figure \ref{fig:overview} summarizes the experimental and analysis workflow. 
We use an established JCA and JSSE benchmark \cite{mousavi_investigation_2024} as a controlled testbed and keep the task description and base Java generation instruction constant when comparing prompt and evidence conditions. 
The design jointly evaluates generated program outcomes and the evidence selected by the automatic retrievers. 

\subsection{Benchmark Tasks and Generator Settings} \label{sec:Benchmark Tasks and Generator Settings}

\input{table_tasks_description}
\input{table_misuse_definitions}

We use 12 JCA and JSSE security functionalities introduced by Mousavi et al. \cite{mousavi_investigation_2024}.
Each functionality has three task wording variants, resulting in 36 task descriptions. 
We collect 30 generations for every model, condition, and task description. 
Therefore, each model--condition setting  contains 1,080 attempts, and the complete study contains 23,760 programs. 
Table~\ref{tab:security-functionalities} shows the task security functionalities and their descriptions \cite{mousavi_investigation_2024}. 
Table~\ref{tab:misuse-definitions} summarizes the misuse categories evaluated in these tasks. 

We evaluate Llama-3.3-70B-Instruct using its official chat template without quantization\footnote{\url{https://huggingface.co/meta-llama/Llama-3.3-70B-Instruct}}, with \texttt{max new tokens=4096}, \texttt{temperature=0.4}, \texttt{top-p=0.8} \cite{arora_optimizing_2024}, and seeds $1234+s$ for sample indices $s=0,1,...,29$, reused for each task-condition pair. 
We access GPT-5.5 through the OpenAI Responses API using the \texttt{gpt-5.5-2026-04-23} snapshot\footnote{\url{https://developers.openai.com/api/docs/models/gpt-5.5}} and \texttt{reasoning\_effort=none}.
In this API configuration, temperature, top-$p$, and random seed were not user-configurable, so we retain the default generation behavior.

\subsection{Security Evidence Corpus and Experimental Conditions} \label{sec:Security Evidence Corpus and Experimental Conditions}

\paragraph{Evidence corpus}
We construct a heterogeneous corpus from publicly available materials commonly consulted by developers when using Java security APIs.
The corpus contains five source tiers: 
(1) API documentation, including the official Oracle JDK API documentation \footnote{\url{https://docs.oracle.com/en/java/javase/21/docs/api/index.html}} and Java Security Standard Algorithm Names \footnote{\url{https://docs.oracle.com/en/java/javase/21/docs/specs/security/standard-names.html}}, where we only include the security APIs-related packages and classes;
(2) Official developer guidance, including the Oracle Security Developer's Guide \footnote{\url{https://docs.oracle.com/en/java/javase/21/security/index.html}} and cryptographic-algorithm configuration guidance\footnote{\url{https://www.java.com/en/configure_crypto.html}};
(3) OWASP guidance, including OWASP Developer Guide Principles of security\footnote{\url{https://devguide.owasp.org/en/02-foundations/03-security-principles/}} and OWASP Developer Guide Principles of cryptography\footnote{\url{https://devguide.owasp.org/en/02-foundations/04-crypto-principles/}};
(4) Secure code examples collected from official samples\footnote{\url{https://docs.oracle.com/javase/8/docs/technotes/guides/security/jsse/samples/}} and developer tutorials \footnote{\url{https://foojay.io/today/securing-symmetric-encryption-algorithms-in-java/}, \url{https://cheatsheetseries.owasp.org/cheatsheets/Java_Security_Cheat_Sheet.html}, \url{https://mkyong.com/java/}, \url{https://dev.java/learn/security/intro/}}; and
(5) Misuse-oriented evidence derived from Mousavi et al. \cite{mousavi_investigation_2024}.
The code examples were not written for the benchmark tasks, and we screen them with CryptoGuard \cite{rahaman_cryptoguard_2019} and manually review them to exclude known security misuse.
HTML sources are split by second-level headings and further divided near 500 words at paragraph boundaries, with approximately 10\% adjacent overlap \cite{khalid_passage_2008}. 
Each complete runnable code example is retained as one chunk, and each misuse pattern is represented as one concise natural language chunk explaining the pattern, its security consequence, and the corresponding avoidance rule. 
The corpus contains 2,954 chunks: 1,269 API documentation items, 1,645 official guidance items, 7 OWASP item descriptions, 20 code examples, and 13 misuse patterns. 

\paragraph{Retrievers}
We use the task description as the retrieval query, keeping queries fixed across retrieval conditions, and compare one sparse retriever (BM25) \cite{robertson_probabilistic_2009} with three dense retrievers.
For the dense retriever, we use FAISS library \cite{douze_faiss_2025} with three different embeddings: jina-embeddings-v2-base-code (Jina embedding) from Jina AI downloaded from Hugging Face\footnote{\url{https://huggingface.co/jinaai/jina-embeddings-v2-base-code}}, gte-Qwen2-7B-instruct (Qwen embedding) from Alibaba downloaded from Hugging Face\footnote{\url{https://huggingface.co/Alibaba-NLP/gte-Qwen2-7B-instruct}}, and text-embedding-3-large (OpenAI embedding) from OpenAI via API\footnote{\url{https://developers.openai.com/api/docs/models/text-embedding-3-large}}. 
Each automatic condition retrieves the top-5 distinct chunks \cite{ke_bridging_2024, wang_coderag-bench_2025}, which are inserted after the task description and before the Java generation instruction.

\paragraph{Prompt and evidence conditions}
We use 11 different conditions:
\begin{enumerate} [leftmargin=*, nosep]
    \item \textbf{Task only baseline:} the task description and a fixed Java generation instruction, without any explicit security requirement. 
    \item \textbf{Security instruction:} the task-only prompt augmented with the instruction: `` You are a senior software engineer. You must always produce secure, modern code using the standard APIs. Avoid insecure or outdated algorithms and modes." It is supplied as a system message for Llama and through the Responses API's instructions field for GPT-5.5.
    \item \textbf{Full-corpus retrieval:} four conditions using BM25 and dense retrievers with the Jina, Qwen, or OpenAI embedding over the complete corpus. 
    \item \textbf{Manual reference evidence:} task-oriented chunks manually selected by the first author from the corpus for each task functionality. This is designed to be a low-noise reference condition, but not an oracle or guaranteed upper bound. 
    \item \textbf{Source-restricted retrieval:} four conditions that apply the same Qwen dense retriever separately to API documentation, official developer guides, code examples, and misuse patterns. 
\end{enumerate}
The Qwen retriever is fixed across the source-restricted conditions to avoid confounding source restriction with retriever variation. 
The OWASP guide is available in the full corpus but is not evaluated as a standalone source-restricted condition its seven chunks provide little scope for selective top-$5$ retrieval.
All conditions share the base instruction: ``Start answer with \texttt{\textasciigrave\textasciigrave\textasciigrave java} on its own line, then write the complete Java program with a main function to test in one single class. Import all used packages and implement all functions. Make sure the program can be compiled using JDK 8. End the program with \texttt{\textasciigrave\textasciigrave\textasciigrave} on its own line.''

\subsection{Data Processing and Evaluation} \label{sec: Data Processing and Evaluation} 
We extract the Java program from the \texttt{\textasciigrave\textasciigrave\textasciigrave java} fence \cite{macedo_output_2025} and evaluate it in two stages. 
First, we check whether the program uses the API required by the task, and whether the program compiles using \texttt{javac} under JDK~8.
JDK~8 is used to preserve compatibility with the CryptoGuard-based evaluation \cite{rahaman_cryptoguard_2019}, and no observed compilation failure was attributable to the JDK versions.
Following the benchmark protocol \cite{mousavi_investigation_2024}, we permit only limited normalization for missing imports and exception handling. 
We do not change any selected algorithm, security parameters, control flow, target API, or other program logic when processing the model outputs. 
An output is classified as \emph{valid} when it uses the required target API and compiles with limited normalization. 
Outputs that omit the target API or fail compilation are regarded as \emph{invalid}.
Second, valid programs are analyzed using CryptoGuard \cite{rahaman_cryptoguard_2019} and are all manually reviewed by the first author using the misuse criteria of Mousavi et al. \cite{mousavi_investigation_2024}. 
The initial manual assessment was made without seeing CryptoGuard findings. 
The first author then compared both assessments and re-examined all disagreements to resolve false positives and false negatives. 
A program is labeled as \emph{detected misuse} if at least one security API misuse is detected. 
Conversely, \emph{no detected misuse} means that none of the security API misuse patterns was identified. 

Let $N$ be all generated attempts, $V$ be valid programs, and $M$ be valid programs with detected misuse. 
We report validity ($V/N$), conditional misuse ($M/V$), and joint yield ($(V-M)/N$).
These metrics capture task completion under the benchmark criterion, detected misuse among valid programs, and the overall production of valid programs with no detected misuse, respectively. 

\subsection{Retrieval Profile Analysis}
Motivated by prior evidence that retrieval quality and generator utility are distinct components of a RAG system \cite{wang_coderag-bench_2025, ke_bridging_2024}, we also analyze the ranked retrieval traces. 
After verifying equality across generator models and repeated samples, the eight automatic conditions yield $8 \times 36=288$ unique query and condition pairs, and $288 \times 5 = 1{,}440$ ranked retrieval chunks. 

For each functionality, the first author constructed a curated task-focused reference set from the corpus that is also used in the \texttt{manual reference evidence} condition. 
This process chooses documentation for required APIs, guidance found through API or misuse searches, functionally similar code examples, and misuse patterns applicable to the task. 
Sets were fixed before the generation experiments and were kept unchanged. 
We evaluate whether automatically retrieved chunks match this set.
For a source-restricted condition, we filter the mapped set to the corresponding evidence tier. 
We compute Mapped Hit@5, which is the fraction of task descriptions retrieving at least one mapped chunk, and Mapped MRR@5, which is the mean reciprocal rank of the first mapped chunk. 
In addition, we measure the proportion of top-$5$ chunks retrieved from each evidence type and the pairwise Jaccard overlap between the four full-corpus retrievers to quantify their similarity.

\begin{figure}[t]
    \centering
    \includegraphics[width=0.5\textwidth]
    {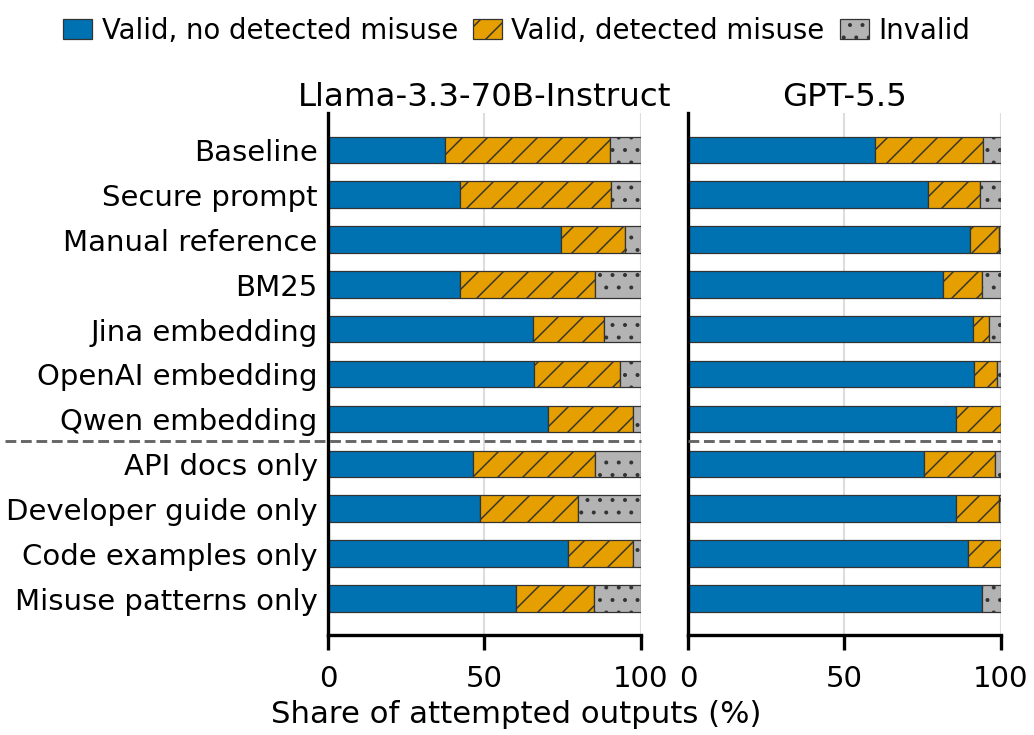}
    \caption{Overall outcome composition across the evaluated model condition pairs. Each bar partitions 1,080 generations into valid programs with no detected misuse, valid programs with detected misuse, and invalid programs.}
    \label{fig:outcome-composition}
\end{figure}

\subsection{Statistical Analysis} \label{sec:Statistical analysis}
The primary security endpoint in this paper is conditional detected misuse $M/V$.
For the analysis of RQ2 results, each prompt-only, full-corpus, and manual reference condition is compared with the task-only baseline. 
For RQ3, each source-restricted condition is compared with full-corpus Qwen embedding retrieval. 
Effects are reported as absolute percentage point differences, with negative values favorable for conditional misuse.

For each experiment condition, one task functionality has three wordings, and we generate 30 samples for each wording. 
Therefore, we estimate uncertainty for the main intervention effects using a nested cluster bootstrap for multilevel data \cite{field_bootstrap_2008}.
Each of 10{,}000 replicates samples 12 task functionalities with replacement and then samples three task descriptions with replacement within every selected functionality. 
All 30 generated samples for a selected wording are kept together. 
The same resampled task descriptions are used across all conditions and both models, and we use the 2.5th and 97.5th percentiles of the bootstrap distribution as the 95\% interval. 
For retriever profile analysis, we use a 10{,}000-replicate functionality-cluster bootstrap that resamples the 12 functionalities and retains their three task description variants, because we use the same manual reference set for all three task description variants within one task functionality.
Using the same resampled task cells, we also compute direct cross-model differences between condition effects to assess whether the observed model-specific result remain stable under task resampling.

\section{Results} \label{sec:Experiment Result}
Figure~\ref{fig:outcome-composition} provides an overview of the three outcomes across all experimental conditions.

\subsection{RQ1: Security API Misuse and Task Heterogeneity} \label{sec: RQ1: Security API Misuse and Task Heterogeneity}

As shown in Fig. \ref{fig:outcome-composition}, security API misuse is common under the task-only baseline prompting.
The Llama-3.3-70B-Instruct model generates 90.1\% (973/1080) valid programs, but 58.5\% (569/973) of them contain at least one detected misuse. 
Only 37.4\% (404/1080) of all generated programs are both valid and free of detected misuse. 
The GPT-5.5 model produces 94.4\% (1019/1080) valid programs, of which 36.6\% (373/1019) contain detected misuse. 
Its corresponding joint yield is 59.8\% (646/1080).
Therefore, the GPT-5.5 model has lower detected misuse in baseline setting, but more than one third of its valid programs still contain security API misuse. 

\begin{figure}[t]
    \centering
    \includegraphics[width=0.5\textwidth]{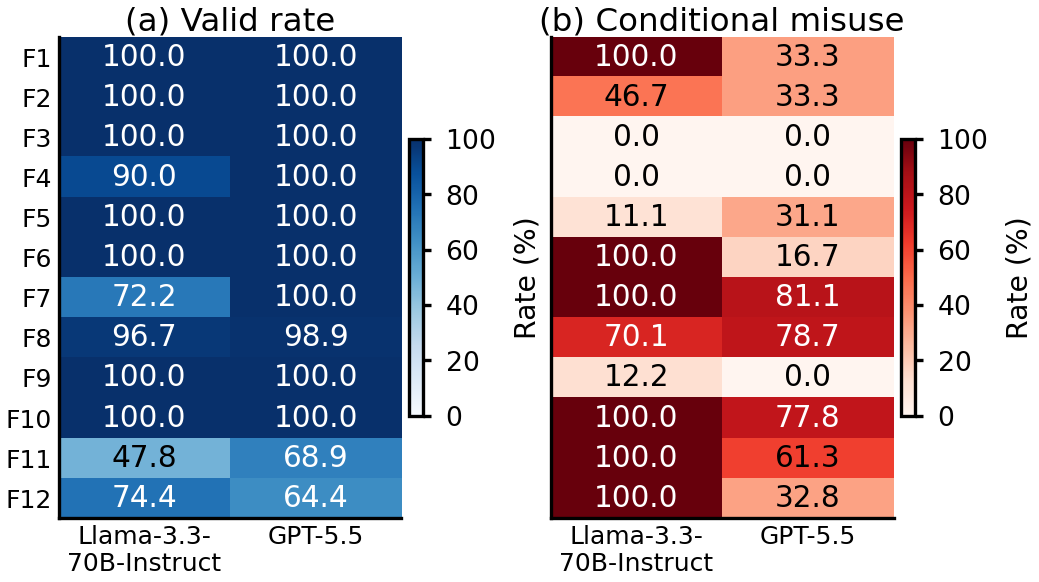}
    \caption{Validity and conditional misuse across the 12 functionalities under the task-only baseline condition.}
    \label{fig:baseline-functionality}
\end{figure}

Figure \ref{fig:baseline-functionality} shows strong functionality level heterogeneity.
For the Llama-3.3-70B-Instruct model, every valid program contains detected misuse in symmetric encryption (F1), message authentication code (F6), key derivation (F7), SSL sockets (F10), hostname verification (F11), and certificate validation (F12).
No misuse is detected for asymmetric encryption (F3) or digital signature (F4).
Validity also varies sharply as it is 100\% for some cryptographic tasks but falls to 47.8\% for hostname verification (F11) tasks.
GPT-5.5 similarly produces no detected misuse for F3, F4, or pseudo random number generator (F9), but substantial risks remain in key derivation, key storage, and SSL socket tasks. 
Its validity is 68.9\% for F11 and 64.4\% for F12, showing that some secure communication tasks remain difficult. 

\noindent\textbf{Answer to RQ1.}
Detected Java security API misuse persists in both evaluated models and is not uniformly distributed. 
The main failures are concentrated in task functionalities that involve secret handling, parameter selection, or multiple verification requirements.

\subsection{RQ2: Effects of Prompting and Full-Corpus Retrieval} \label{RQ2: Effects of Prompting and Full-Corpus Retrieval}

\begin{figure}[t]
    \centering
    \includegraphics[width=0.5\textwidth]
    {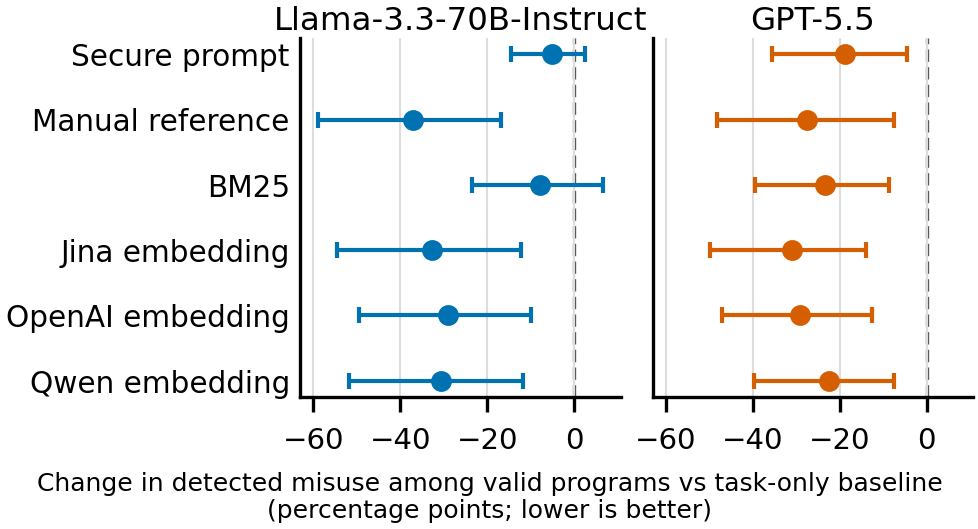}
    \caption{Effects of the interventions on detected misuse among valid programs. Points denote absolute percentage point changes relative to the task-only baseline.
    Error bars are 95\% percentile intervals from 10,000 nested cluster bootstrap replicates.
    Negative values indicate lower detected misuse.}
    \label{fig:rq2-effects}
\end{figure}

Fig. \ref{fig:rq2-effects} shows the changes in conditional misuse relative to the task-only baseline. 
For the Llama-3.3-70B-Instruct model, the security instruction reduces the observed rate by 5.2 percentage points, while BM25 retriever reduces it by 7.9 percentage points. 
The three dense retrievers produce larger reductions: 32.7 percentage points for the Jina embedding, 30.7 for the Qwen embedding, and 29.0 for the OpenAI embedding. 
Manual reference evidence leads to the largest reduction, which is 37.0 percentage points. 
For the GPT-5.5 model, all six intervention intervals exclude zero. 
The security instruction reduces conditional misuse by 18.9 percentage points, and the corresponding reductions are 23.4 percentage points for BM25 retriever, 31.0 for the Jina embedding, 22.4 for the Qwen embedding, 29.2 for the OpenAI embedding, and 27.5 for manual reference evidence. 
For GPT-5.5, the Jina embedding leads to the lowest conditional misuse among all full-corpus retrievers (5.6\%), while the OpenAI embedding produces the highest joint yield (91.4\%). 
The Qwen embedding leads to 100\% validity, but it has a higher conditional misuse rate at 14.2\%. 

The model-specific results differ, particularly for the prompt-only and BM25 retriever. 
The manual reference evidence is competitive for both models, but it does not descriptively dominate every automatic retriever for the GPT-5.5 model. 

\noindent\textbf{Answer to RQ2.}
Full-corpus dense retrieval consistently leads to large reductions in detected misuse, but the prompt-only and sparse retrieval effects are less stable for the Llama-3.3-70B-Instruct model. 
Retrieval choice also affects whether the gain appears primarily as lower misuse, higher validity, or more valid programs with no detected misuse.

\subsection{RQ3: Retrieval Profiles and Source Restrictions}\label{sec: RQ3: Retrieval Profiles and Source Restrictions}

\subsubsection{Full-Corpus Retrieval Profiles}

\begin{figure}[t]
    \centering
    \includegraphics[width=0.5\textwidth]
    {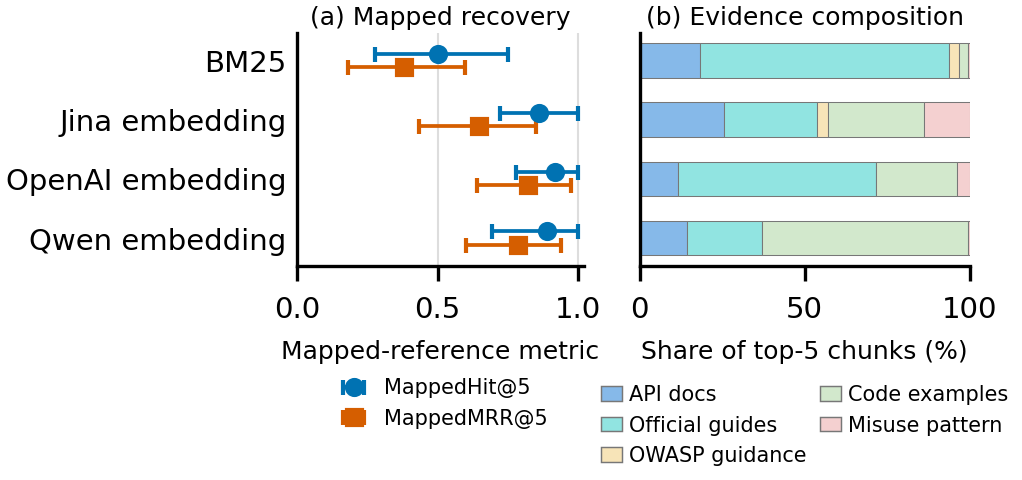}
    \caption{Full-corpus retrieval profiles.
    (a) Mapped-reference Hit@5 and MRR@5 (hit fraction and mean reciprocal rank), with 95\% functionality-cluster bootstrap intervals.
    (b) Evidence-source composition over the 180 top-$5$ chunks retrieved by each retriever.}
    \label{fig:retrieval-profiles}
\end{figure}

Fig. \ref{fig:retrieval-profiles} shows the evidence selected by the four full-corpus retrievers. 
BM25 retrieves at least one mapped reference for 50.0\% of the 36 task descriptions and achieves a mapped MRR@5 of 0.380.
The corresponding Mapped Hit@5 and MRR@5 values are 86.1\% and 0.649 for the Jina embedding, 91.7\% and 0.819 for the OpenAI embedding, and 88.9\% and 0.787 for the Qwen embedding. 
The evidence composition differs substantially across retrievers. 
BM25 retrieves about 75.6\% official developer guidance chunks in its top-$5$ retrievals. 
The OpenAI embedding dense retriever retrieves primarily official developer guidance (60.0\%), followed by code examples (24.4\%).
However, the Qwen embedding retrieval result is dominated by code examples (62.2\%), while the Jina embedding produces the most heterogeneous profile, where it includes 25.6\% API documentation, 28.3\% official developer guidance, 3.3\% OWASP material, 28.9\% code examples, and 13.9\% misuse patterns. 
The retrievers also select very different evidence chunks. 
Their mean pairwise Jaccard@5 ranges from 0.071 (BM25 and the Jina embedding) to 0.283 (the Qwen embedding and OpenAI embedding), and there are no retriever pairs that produce the identical top-$5$ retrieved set for any task description.

Together, the recovery and composition results show two complementary dimensions of retrieval behavior. 
The OpenAI embedding dense retriever most consistently recovers the task-focused mapped reference, while the Jina embedding provides LLMs with more diverse evidence types and has the lowest conditional misuse for both models among full-corpus retrievers. 
Therefore, retriever choice affects both which task-relevant evidence can be recovered and how different evidence roles can be combined in the augmented generation context.

\subsubsection{Source-Restricted Retrieval}

Table \ref{tab:source-restriction} compares the four source-restricted conditions with the full-corpus Qwen embedding retrieval. 
For the Llama-3.3-70B-Instruct model, code examples retrieval has the most favorable result with a 21.4\% conditional misuse, and 97.5\% validity, and achieves a $-6.4$ percentage point conditional misuse change relative to the full Qwen embedding. 
Restricting retrieval to API documentation or developer guidance produces lower validity and joint yield. 
Misuse patterns only retrieval also reduces validity to 85.1\% and keeps the conditional misuse relatively stable compared with full Qwen. 
The GPT-5.5 model profile differs from the Llama-3.3-70B-Instruct model. 
Code examples only retrieval maintains 100\% validity and reduces conditional misuse to 10.5\%.
Misuse patterns only retrieval produces no detected misuse among 1,016 valid programs, but it reduces validity from 100\% under the full Qwen embedding retrieval to 94.1\%.

The observed source-restricted profiles differ between model settings. 
The cross model interaction (GPT-5.5 effect minus Llama-3.3-70B-Instruct effect) for misuse patterns is $-15.9$ percentage points with a 95\% interval of $[-40.0,\,5.3]$.
The corresponding differences and 95\% intervals are $-9.0$ percentage points and $[-33.2,\,14.3]$ for API documentation, $-11.7$ percentage points and $[-37.9,\,15.7]$ for official developer guidance, and  $2.7$ percentage points and $[-15.8,\,20.4]$ for secure code examples. 
All of the cross-model conditional misuse interaction intervals include zero, which shows the differences are descriptive but not always stable. 

\noindent\textbf{Answer to RQ3.}
Different full-corpus retrievers provide different mixtures of security evidence to models, and the source-restricted retrieval conditions suggest complementary roles for evidence sources. 
Code examples yield the lowest observed restricted-source conditional misuse for Llama-3.3-70B-Instruct, while explicit misuse patterns yield the lowest observed conditional misuse for GPT-5.5, with a corresponding validity trade off.

\begin{table*}[t]
\centering
\footnotesize
\setlength{\tabcolsep}{3.5pt}
\caption{Outcomes under full-corpus Qwen retrieval and the four source-restricted conditions. 
V/M/J reports validity, detected misuse among valid programs, and joint yield, respectively, in percent. 
$\Delta M$ is the absolute change in conditional misuse relative to full-corpus
Qwen retrieval; lower values are favorable. 
Intervals are 95\% nested cluster bootstrap intervals.}
\label{tab:source-restriction}
\begin{tabular}{lcc cc}
\toprule
& \multicolumn{2}{c}{Llama-3.3-70B-Instruct}
& \multicolumn{2}{c}{GPT-5.5} \\
Condition
& V/M/J (\%)
& $\Delta M$ pp [95\% CI]
& V/M/J (\%)
& $\Delta M$ pp [95\% CI] \\
\midrule
Full Qwen
& 97.3/27.8/70.3 & --
& 100.0/14.2/85.8 & -- \\

API docs
& 85.3/45.7/46.3 & +17.9 [-3.3, 42.0]
& 98.2/23.1/75.6 & +8.9 [-0.7, 20.8] \\

Developer guide
& 79.9/39.4/48.4 & +11.6 [-10.3, 35.5]
& 99.5/14.0/85.6 & -0.1 [-9.7, 9.6] \\

Code examples
& 97.5/21.4/76.7 & -6.4 [-22.0, 7.5]
& 100.0/10.5/89.5 & -3.7 [-12.7, 2.2] \\

Misuse patterns
& 85.1/29.5/60.0 & +1.7 [-17.5, 22.1]
& 94.1/0.0/94.1 & -14.2 [-30.6, -1.6] \\
\bottomrule
\end{tabular}
\end{table*}

\section{Discussion} \label{sec:Discussion}

\subsection{Residual Misuse as a Failure of Constraint Completion} \label{sec:Residual Misuse as a Failure of Constraint Completion}
The baseline results align with prior studies showing that plausible, executable LLM-generated code does not necessarily imply correctness or security \cite{perry_users_2023, pearce_asleep_2025, khoury_how_2023, he_large_2023}.
The remaining difficulty is not simply whether an LLM knows a Java security API.
High-risk functionalities require several security-relevant decisions that might not be visible in the functional task description.
For example, key derivation and key storage require appropriate secret handling, salt and iteration choices, and password life cycle decisions. 
In addition to establishing a connection, secure communication tasks also require hostname or certificate verification. 
Prior studies of security API usability show that developers similarly struggle to coordinate multiple
cryptographic-API calls and security constraints \cite{nadi_jumping_2016, kruger_crysl_2021}.

We interpret this pattern as consistent with incomplete security constraint specification.
An underspecified coding task typically states the primary functional goal while leaving algorithm choices, credential handling, and verification requirements implicit \cite{sajadi_llms_2025, assal_security_2018}.
An LLM may generate a plausible program but the omitted details result in unsafe patterns. 
The large gains from dense retrievers show that explicit evidence can mitigate this failure, but residual misuse suggests that providing relevant knowledge alone does not guarantee that every constraint is satisfied.

\subsection{Evidence Form Shapes How Retrieval Helps} \label{sec: Evidence Form Shapes How Retrieval Helps}

The security evidence is not a homogeneous resource. 
API documentation supplies interface facts, developer guidance expresses normative requirements, code examples provide complete implementation paths, and misuse patterns show explicit negative constraints. 
The full-corpus retrievals combine these sources in different proportions, which creates different contexts for the models. 

The manual reference condition confirms that task-focused curation is effective, while automatic retrieval shows that several useful evidence combinations can support the same task. 
The OpenAI embedding dense retriever most consistently recovers the curated reference, while the Jina embedding dense retriever constructs the most heterogeneous selection and produces the lowest observed conditional misuse among full-corpus retrievers, and the Qwen embedding dense retriever strongly favors code examples. 
These results are consistent with prior work distinguishing retrieval from the model's capability to apply retrieved context \cite{wang_coderag-bench_2025, ke_bridging_2024}.

The source-restricted results further clarify the roles of different evidence forms. 
For the Llama-3.3-70B-Instruct model, code examples yield the lowest observed restricted-source conditional misuse, consistent with prior findings that executable demonstrations can be more directly useful than descriptive text \cite{chen_when_2025, zhang_seccoder_2024}. 
For the GPT-5.5 model, task-aligned misuse patterns are highly effective, which shows that explicit negative constraints can be actionable in this evaluated setting. 
However, the condition also reduces validity, showing that avoiding known misuse patterns does not by itself ensure complete task completion. 
Because the candidate corpora differ in size and specificity and the cross-model intervals include zero, these findings cannot be interpreted as a universal source hierarchy.

\subsection{Implications for Security-Aware Coding Assistant} \label{sec: Implications for Security-Aware Coding Assistant Systems}

\paragraph{Retrieve for coverage and actionability}
Security-oriented retrieval should cover the implementation's security decisions, rather than optimize similarity alone. 
This may require API documentation, normative guidance, complete examples, and misuse warnings. 
Source-aware or hybrid retrieval and security-oriented re-ranking could combine these evidence roles. 
Their balance should be validated through end-to-end evaluation on the intended model and task distribution. 

\paragraph{Use validation feedback to repair or prevent misuse}
Avoiding misuse by producing uncompilable code or omitting the required API does not complete the task. 
Coding assistants should monitor validity and security API misuse together. 
When no generation satisfies both functionality requirements and security checks, a coding assistant should not present the output as complete to the user. 
Compilation and security checks should trigger an explicit warning, repair, or regeneration. 
This turns validity and misuse detection from an evaluation metric into operational controls in the generation workflow.

\paragraph{Retain post-generation security validation}
Models may still omit security constraints despite receiving relevant evidence. 
Therefore, post-generation static analysis, security check rules, and manual code review remain necessary deployment controls. 
In addition, the coding assistant system can combine the RAG system with state of the art decoding methods from the perspective of instruction tuning or constrained decoding \cite{mukherjee_sosecure_2025, shi_rescue_2026, xu_prosec_2025, zhang_seccoder_2024}.
As a result, the benefit of external knowledge can be more stable and the overall security of LLM generated code can be further improved.

\section{Threats to Validity} \label{sec: Threats to Validity}

\paragraph{Construct validity}
The outcome definitions capture only part of secure code generation. 
A program is valid when it uses the required target API and compiles after limited normalization for missing imports or checked exception handling. 
But this does not guarantee an intended runtime behavior or deployment suitability. 
Similarly, ``no detected misuse'' means that none of the misuse patterns listed in \ref{tab:misuse-definitions} \cite{mousavi_investigation_2024} were identified, but this does not mean the program is free of vulnerabilities involving other APIs or application-level weaknesses.
The retrieval metrics also have limitations. 
The mapped reference sets are manually curated task-oriented low noise references, but are not an oracle. 
An unmapped retrieved chunk may not be irrelevant, and the mapped Hit@5 and MRR@5 only measures the recovery of the curated knowledge set. 
The misuse patterns corpus is derived from the same known misuse categories that are used in evaluation. 
Therefore, its strong performance measures only the usefulness of task-aligned negative constraints for avoiding known patterns, but cannot show generalization to unseen vulnerabilities. 

\paragraph{Internal and statistical validity}
Reference curation and program review were performed by one author, so independent annotation agreement was not measured. 
We use fixed task descriptions, prompt templates, chunking rules, and generation instruction to make experiment conditions comparable. 
However, alternative prompt formulations, context ordering, chunk sizes, or the numbers of chunks being retrieved may change how the retrieved evidence is used. 
The source-restricted condition comparisons also change more than the surface representation of knowledge because their candidate corpora differ in size, specificity, redundancy, and retrieval difficulty. 
The four source-restricted conditions share the Qwen embedding dense retriever. 
This controls retriever variation but makes the conclusions conditional on that retriever. 
The model comparisons are also specific to the evaluated inference settings. 
The Llama-3.3-70B-Instruct model uses fixed sampling parameters, while those parameters are not configurable for the GPT-5.5 model. 
The provider-side nondeterminism may cause unstable performance.
In the experiment, each task description is sampled 30 times, but these generations share the same prompt and cannot be treated as 30 independent programming tasks. 
Therefore, the primary uncertainty analysis resamples complete functionality and wording clusters using a nested bootstrap for multilevel data \cite{field_bootstrap_2008}.
The intervals can describe sensitivity to the benchmark, but may not transfer directly to other programming languages, security libraries, or non-API security failures.

\paragraph{External validity}
The study covers 12 JCA and JSSE functionalities implemented as self-contained, single-class Java programs. 
They do not represent repository-level development, concurrent systems, or the full range of software vulnerabilities. 
We evaluate one proprietary model snapshot and one open-weight model configuration, so the observations may not be generalized to all closed or open-weight models.
Our evaluation is also limited to single-turn, one-shot program generation. 
Each model receives a fixed task and evidence prompt and produces one complete Java program without observing compilation errors, execution results, or any feedback. 
Modern coding agents can interact iteratively with repositories and execution environments, run tests and use other tools, and revise outputs based on the resulting feedback \cite{yang_swe_2024, li_codetree_2025}. 
Such loops may improve validity or security. 
Therefore, our findings only characterize first-pass generation, but not the final quality through iterative agentic repair.

\section{Conclusion} \label{sec:Conclusion}

This paper presents an end-to-end, misuse-centered analysis of external security evidence for LLM-generated Java code. 
Across 23,760 generations, JCA and JSSE API misuse remains common under task-only prompting and is concentrated in functionalities that require multiple security constraints to be satisfied at the same time. 
Full-corpus dense retrieval substantially reduces detected misuse, and mapped-reference recovery results show that different retrievers tend to select different types of security knowledge. 
The source-restricted conditions also suggest no universally dominant evidence source, and different models may have different capabilities to utilize different types of knowledge. 
This suggests that security-oriented code RAG should retrieve evidence for coverage and actionability rather than rely on a single source of knowledge. 
Reliable secure coding assistants should combine model-aware knowledge selection, retriever policies, post-generation review, and iterative repair.

\section*{Acknowledgements}
Robert Biddle is affiliated with Carleton University in Ottawa, Canada, and acknowledges the support of the Natural Sciences and Engineering Research Council of Canada (NSERC), RGPIN-2022-04887.

\bibliographystyle{IEEEtran}
\bibliography{references}

\end{document}

%% file: table_tasks_description.tex
\begin{table*}[!t]
\centering
\footnotesize 
\setlength{\tabcolsep}{1pt}
\renewcommand{\arraystretch}{1.0}
\caption{JCA and JSSE task functionalities used in this study and their descriptions.}
\label{tab:security-functionalities}
\begin{tabularx}{\textwidth}{
    >{\centering\arraybackslash}p{0.05\textwidth}
    >{\raggedright\arraybackslash}p{0.3\textwidth}
    >{\raggedright\arraybackslash}X
}
\toprule
\textbf{API} & \textbf{Security Task Functionalities} & \textbf{Description} \\
\midrule

\multirow{9}{*}{\textbf{JCA}}
& F1: Symmetric Encryption
& Uses a single key for encryption and decryption.
\\

& F2: Symmetric Encryption in CBC mode
& Symmetric encryption in Cipher Block Chaining (CBC) mode.
\\

& F3: Asymmetric Encryption
& Uses two keys, a public key for encryption and a private key for decryption.
\\

& F4: Digital Signature
& Sender signs data with private key and receiver verifies it with public key.
\\

& F5: Hash Functions
& Converts input data into hash values to maintain data integrity.
\\

& F6: Message Authentication Code
& Uses hashing and a secret key to generate a code for data integrity and authenticity.
\\

& F7: Key Derivation Function
& Generates a secure cryptographic key from a given password.
\\

& F8: Key Storage
& Securely stores sensitive credentials using a password.
\\

& F9: Pseudo Random Number Generator
& Generates random numbers securely for cryptographic applications.
\\

\midrule

\multirow{3}{*}{\textbf{JSSE}}
& F10: SSL Socket Establishment
& Creates an SSL socket for secure connection between a host and a client.
\\

& F11: Hostname Verification
& Confirms server hostname matches the intended hostname.
\\

& F12: Certificate Validation
& Verifies if the certificate is issued and signed by a trusted certificate authority.
\\

\bottomrule
\end{tabularx}
\end{table*}

%% file: table_misuse_definitions.tex
\begin{table}[!t]
\centering
\footnotesize
\setlength{\tabcolsep}{1pt}
\renewcommand{\arraystretch}{1.0}
\caption{Common API misuses.}
\label{tab:misuse-definitions}
\begin{tabularx}{\linewidth}{
    >{\raggedright\arraybackslash}p{0.2\linewidth}
    >{\raggedright\arraybackslash}X
}
\toprule
\textbf{Misuse Type} & \textbf{Description} \\
\midrule
M1  & Constant or predictable cryptographic keys. \\
M2  & Insecure mode of operation for encryption. \\
M3  & Predictable Initialization Vectors (IVs) in CBC mode. \\
M4  & Short cryptographic keys. \\
M5  & Constant or predictable seeds for PRNG. \\
M6  & Constant or predictable salts for key derivation. \\
M7  & Insufficient number of iterations for key derivation. \\
M8  & Hardcoded constant passwords. \\
M9  & Broken hash functions. \\
M10 & Compromised MAC algorithms. \\
M11 & Improper SSL socket. \\
M12 & Improper hostname verification. \\
M13 & Improper certificate validation. \\
\bottomrule
\end{tabularx}

\end{table}